\documentclass[aps,prd,preprintnumbers,superscriptaddress,nofootinbib,showpacs,twocolumn]{revtex4}%
\usepackage[dvips]{graphicx}
\usepackage{bm,latexsym,amsmath,amssymb,amsfonts,mathrsfs}
\usepackage{color}
\input{colordvi.tex}
\newcommand*{\D}{{\rm d}}
\newcommand*{\mpl}{M_{\rm Pl}}
\begin{document}

\title{Cosmological matching conditions
and galilean genesis in Horndeski's theory}

\author{Sakine~Nishi}
\email[Email: ]{sakine_n"at"rikkyo.ac.jp}
\affiliation{Department of Physics, Rikkyo University, Toshima, Tokyo 175-8501, Japan
}

\author{Tsutomu~Kobayashi}
\email[Email: ]{tsutomu"at"rikkyo.ac.jp}
\affiliation{Department of Physics, Rikkyo University, Toshima, Tokyo 175-8501, Japan
}

\author{Norihiro~Tanahashi}
\email[Email: ]{norihiro.tanahashi"at"ipmu.jp}
\affiliation{Kavli Institute for the Physics and Mathematics of the Universe (IPMU),
The University of Tokyo, Kashiwa, Chiba, 277-8568, Japan
}
\affiliation{Department of Applied Mathematics and Theoretical Physics,
University of Cambridge, Wilberforce Road, Cambridge CB3 0WA, UK
}

\author{Masahide~Yamaguchi}
\email[Email: ]{gucci"at"phys.titech.ac.jp}
\affiliation{Department of Physics, Tokyo Institute of Technology, Tokyo 152-8551, Japan
}

\begin{abstract}
We derive the cosmological matching conditions for the homogeneous and
isotropic background and for linear perturbations in Horndeski's most
general second-order scalar-tensor theory.  In general relativity, the
matching is done in such a way that the extrinsic curvature is
continuous across the transition hypersurface.  This procedure is
generalized so as to incorporate the mixing of scalar and gravity
kinetic terms in the field equations of Horndeski's theory.  Our
matching conditions have a wide range of applications including the
galilean genesis and the bounce scenarios, in which stable, null energy
condition violating solutions play a central role.  We demonstrate how
our matching conditions are used in the galilean genesis scenario. In
doing so, we extend the previous genesis models and provide a unified
description of the theory admitting the solution that starts expanding
from the Minkowski spacetime.
\end{abstract}

\preprint{RUP-14-1, IPMU14-0003}
\maketitle

\section{Introduction}

Scalar fields are ubiquitous in cosmology.
Inflation~\cite{inflation} is considered to be driven by one or multiple scalar fields,
which can seed the large-scale structure of the Universe as well.
The current cosmic acceleration may also be caused by
a scalar field dominating the energy content of the Universe
as dark energy (see e.g.~\cite{d-e} for a review).
A great variety of modified gravity models have been proposed
as an alternative to dark energy (see e.g.~\cite{mog} and references therein),
many of which involve an additional scalar degree of freedom
in the gravity sector.
Early-universe scenarios other than inflation have also been
explored (see e.g.~\cite{Brandenberger:2009jq} for a recent review),
such as bounce models, and they are often
based on some scalar-field theory.

Almost forty years ago, Horndeski constructed
the most general theory composed of
the metric $g_{\mu\nu}$ and the scalar field $\phi$
with second-order field equations~\cite{Horndeski},
which has long been ignored until recently~\cite{fab4}.
In the course of generalizing the galileon scalar-field
theory, Horndeski's theory was rediscovered in
its modern form~\cite{Nicolis:2008in, Deffayet:2009wt, Deffayet:2011gz}.
(The equivalence of the generalized galileon and
Horndeski's theory was first shown in Ref.~\cite{Kobayashi:2011nu}.)
The action is given by
\begin{eqnarray}
S_{\rm Hor}&=&\int\D^4x\sqrt{-g}\left({\cal L}_2
+{\cal L}_3+{\cal L}_4+{\cal L}_5\right),\label{Hor}
\end{eqnarray}
with
\begin{eqnarray}
&&{\cal L}_2=G_2(\phi, X),\quad
{\cal L}_3=-G_3(\phi, X)\Box\phi,
\nonumber\cr&&
{\cal L}_4=G_4(\phi, X)R
+G_{4X}\left[(\Box\phi)^2-(\nabla_\mu\nabla_\nu\phi)^2\right],
\nonumber\cr&&
{\cal L}_5=G_5(\phi, X)G^{\mu\nu}\nabla_\mu\nabla_\nu\phi
-\frac{1}{6}G_{5X}\bigl[(\Box\phi)^3
\nonumber\\&&\qquad\qquad
-3\Box\phi(\nabla_\mu\nabla_\nu\phi)^2
+2(\nabla_\mu\nabla_\nu\phi)^3\bigr],
\end{eqnarray}
where
$X:=-g^{\mu\nu}\partial_\mu\phi\partial_\nu\phi/2$,
$R$ is the Ricci scalar, and $G_{\mu\nu}$ is the Einstein tensor, and
$G_2, G_3, G_4$, and $G_5$ are arbitrary functions of $\phi$ and $X$.
(Here and hereafter we use the notation $G_{iX}:=\partial G_i/\partial X$,
$G_{i\phi}:=\partial G_i/\partial \phi$, and so on.)
Since this theory contains all the single-field inflation models
and modified gravity models with one scalar degree of freedom
as specific cases,
it is of great importance in cosmology and hence
considerable attention has been paid in recent years to
various aspects of Horndeski's theory (see the nonexhaustive list of
references~\cite{CosHor}).

In this paper, we will address the following issue:
suppose that the Universe undergoes a sharp transition
caused, for example, by sudden halt of the scalar field
or by a discontinuous jump in matter pressure, and then
what are the continuous quantities across the transition hypersurface
in Horndeski's theory?
In general relativity,
it is known that the induced metric on the surface and
its extrinsic curvature must be continuous.
This implies that the Hubble parameter $H$ is continuous.
As for linear cosmological perturbations,
the matching conditions in general relativity
are clarified in Refs.~\cite{Deruelle:1995kd, Hwang:1991an}.
In Horndeski's theory, however,
scalar and gravity kinetic terms are mixed
due to second derivatives on $\phi$ in the Lagrangian~\cite{Deffayet:2010qz},
and as a result the matching conditions would be nontrivial both for
the homogeneous and isotropic background and for cosmological perturbations.
This point was raised in the context of galilean genesis~\cite{Creminelli:2010ba}
and was studied based on specific Lagrangians~\cite{Creminelli:2012my,Hinterbichler:2012yn}.
In this paper, we start from the boundary terms in Horndeski's theory~\cite{Padilla:2012ze}
and derive rigorously the cosmological matching conditions
in their most general form.

The matching conditions obtained in this paper
have a wide range of applications. In particular,
Horndeski's theory allows for stable violation of
the null energy condition (NEC), leading to
interesting possibilities such as galilean genesis mentioned above
and non-singular bounce models~\cite{Qiu:2011cy,Easson:2011zy,Cai:2012va,Osipov:2013ssa,Cai:2013kja}.
Our matching conditions provide a generic algorithm to follow the evolution of
the cosmological background and perturbations, which is applicable to those scenarios.

This paper is organized as follows.
In the next section we summarize the boundary terms in Horndeski's theory~\cite{Padilla:2012ze},
which are the basis of the present work.
Then, in Sec.~III, we derive the cosmological matching conditions both for
the background and perturbations.
To present an example, we develop a unified Lagrangian
accommodating all the previous models of galilean genesis,
and apply our matching conditions to this general model in Sec.~IV.
We draw our conclusions in Sec.~V.

\section{Boundary terms for Horndeski's theory}

We begin with summarizing the result of Ref.~\cite{Padilla:2012ze}.
The action we are going to study is given by
\begin{eqnarray}
S=S_{\rm Hor}+S_{\rm m}+S_{\rm B},\label{total-action}
\end{eqnarray}
where $S_{\rm Hor}$
is Horndeski's action~(\ref{Hor}),
$S_{\rm m}$ is the action for usual matter,
and $S_{\rm B}$ is the boundary term.
This last term is necessary
when one considers a spacetime ${\cal M}$ divided into two domains, ${\cal M}_\pm$,
by a surface $\Sigma$.
In what follows $\Sigma$ is supposed to be spacelike.

Let us take a look at the case of general relativity.
The variation of the Einstein-Hilbert term with respect to the metric involves
a normal derivative of the metric variation,
\begin{eqnarray}
\delta_g\left(\int\D^4x\sqrt{-g}R\right)
\supset
-\int_{\Sigma_\pm}\D^3 x\sqrt{\gamma}\gamma^{\mu\nu }
n^\lambda\nabla_\lambda\delta g_{\mu\nu},\label{mveh}
\end{eqnarray}
where $\gamma_{\mu\nu}=g_{\mu\nu}+n_\mu n_\nu$ is the induced metric on $\Sigma_\pm$
and $n^\mu$ is the future directed unit normal.
Here and hereafter we write
\begin{eqnarray}
\int_{\Sigma_{\pm}}(\cdots):=\int_{\Sigma_+}(\cdots)-\int_{\Sigma_-}(\cdots),
\end{eqnarray}
with $\Sigma_\pm$ denoting the two sides of $\Sigma$. The presence of
the normal derivative of the metric variation is problematic;
to obtain a well-defined variational problem,
one has to add a boundary term that cancels the contribution~(\ref{mveh}).
By noticing that the variation of the trace of the extrinsic curvature,
$K_{\mu\nu}:=\gamma_{\mu}^{\;a}\gamma_{\nu}^{\;b}\nabla_{(a}n_{b)}$,
gives rise to the same contribution,
\begin{eqnarray}
\delta_g\left(\sqrt{\gamma} K\right)
\supset \frac{1}{2}\sqrt{\gamma}\gamma^{\mu\nu}n^\lambda\nabla_\lambda\delta g_{\mu\nu},
\end{eqnarray}
we are lead to add the well-known Gibbons-Hawking term on the boundary~\cite{GHterm}.

Since the most general scalar-tensor Lagrangian having second-order field equations
contains second derivatives of the scalar field which is nonminimally coupled to gravity,
as well as the second derivatives of the metric,
the corresponding boundary action is not simply given by the Gibbons-Hawking term.
For example,
$G_3\Box\phi$ produces the following problematic normal derivative:
\begin{eqnarray}
\delta_\phi\left(\int\D^4 x\sqrt{-g} G_3\Box\phi\right)
\supset \int_{\Sigma_\pm}\D^3 x\sqrt{\gamma}G_3n^\mu\nabla_\mu\delta\phi.
\end{eqnarray}
This can be canceled by adding
\begin{eqnarray}
B_3= \int_{\Sigma_{\pm}}\D^3x\sqrt{\gamma} F_3,
\end{eqnarray}
where
\begin{eqnarray}
F_3(\phi, X_0, \widetilde X):=\int^{X_0}_0\frac{\D u}{\sqrt{2u}}G_3(\phi, u+\widetilde X),
\end{eqnarray}
with $X_0:=(n^\mu\nabla_\mu\phi)^2/2$ and
$\widetilde X:=-\gamma^{\mu\nu}\partial_\mu\phi\partial_\nu\phi/2$.
(Note that $X=X_0+\widetilde X$.)
Similarly, one can obtain the boundary contributions
corresponding to ${\cal L}_4$ and ${\cal L}_5$.
The boundary term for the galileon Lagrangian
was considered in Ref.~\cite{Dyer:2009yg}, and then
the complete boundary term in Horndeski's theory,
which is composed of three different parts,
$S_{\rm B}=B_3+B_4+B_5$,
was derived for the first time in
Ref.~\cite{Padilla:2012ze}. The latter two terms are given by
\begin{eqnarray}
B_4&=&2\int_{\Sigma_\pm}\D^3x\sqrt{\gamma}\left(G_4K-F_{4\widetilde X}D^2\phi\right),
\\
B_5&=&\int_{\Sigma_\pm}\D^3 x\sqrt{\gamma}\biggr\{
\frac{1}{2}G_5\left(K^2-K_{\mu\nu}K^{\mu\nu}\right)
n^\lambda\nabla_\lambda\phi
\nonumber\\&&\quad
-G_5\left( KD^2\phi - K^{\mu\nu}D_{\mu}D_\nu\phi\right)+\frac{1}{2}F_5 R^{(3)}
\nonumber\\&&\quad
+\frac{1}{2}F_{5\widetilde X}\left[(D^2\phi)^2-D_\mu D_\nu\phi D^\mu D^\nu\phi\right]\biggl\},
\end{eqnarray}
where each $F_i\;(i=4,5)$ is defined similarly to $F_3$ as
\begin{eqnarray}
F_i(\phi, X_0, \widetilde X):=\int^{X_0}_0\frac{\D u}{\sqrt{2u}}G_i(\phi, u+\widetilde X),
\end{eqnarray}
$D_\mu$ is the covariant derivative on the boundary, and $R^{(3)}$ is
the boundary Ricci scalar.

Having found the boundary term,
one can obtain the junction conditions
that describe discontinuity across the hypersurface $\Sigma$,
as a generalization of Israel's conditions~\cite{Israel}.
The variational principle for~(\ref{total-action}) yields the equations of motion and
\begin{eqnarray}
\delta S\supset \int_{\Sigma_\pm}\D^3x\sqrt{\gamma}
\left({\cal J}^{\mu\nu}\delta\gamma_{\mu\nu}+{\cal J}^\phi\delta\phi\right).
\label{delS}
\end{eqnarray}
Here,
${\cal J}^{\mu\nu}={\cal J}^{\mu\nu}_3+{\cal J}^{\mu\nu}_4+{\cal J}^{\mu\nu}_5$,
with\footnote{In
deriving Eq.~(\ref{Jmn3}) we used
\begin{eqnarray*}
F_3&=&\int_0^{X_0}\!\!\!\D u\,\partial_u\left(\sqrt{2u}\right)G_3(\phi, u+\widetilde X)
\\
&=& G_3(\phi, X)n^\mu\nabla_\mu\phi-\int^{X_0}_0
\D u\sqrt{2u}\,G_{3u}(\phi, u+\widetilde X),
\end{eqnarray*}
to rearrange the original expression of Ref.~\cite{Padilla:2012ze}.
}
\begin{eqnarray}
{\cal J}^{\mu\nu}_3&=&-\frac{1}{2}\gamma^{\mu\nu}\int^{X_0}_0
\D u\sqrt{2u}\,G_{3u}(\phi, u+\widetilde X)
\nonumber\\&&
+\frac{1}{2}F_{3\widetilde X}D^\mu\phi
D^\nu\phi,\label{Jmn3}
\end{eqnarray}
and lengthy expressions for ${\cal J}^{\mu\nu}_4$ and ${\cal J}^{\mu\nu}_5$,
for which we refer the reader to
Ref.~\cite{Padilla:2012ze}.\footnote{In arXiv:1206.1258v1~\cite{Padilla:2012ze}
there is a typo in the expression for ${\cal J}^{\mu\nu}_5$, so
the reader should refer to the updated version of Ref.~\cite{Padilla:2012ze}.}
In the case of general relativity ($G_4=$ const, $G_3=0=G_5$),
one finds ${\cal J}_{\mu\nu}=-G_4(K_{\mu\nu}-\gamma_{\mu\nu}K)$.
A concrete expression for ${\cal J}^\phi$ is also found in Ref.~\cite{Padilla:2012ze}.

We allow for a localized source on $\Sigma$ whose
action is denoted by $S_{\Sigma}$.
Variation of the action $S_\Sigma$ will take the form
\begin{eqnarray}
\delta S_\Sigma = \int_\Sigma
\D^3x\sqrt{\gamma}\left(\frac{1}{2}\tau^{\mu\nu}\delta\gamma_{\mu\nu}
-\Delta^\phi\delta\phi\right),\label{delSSigma}
\end{eqnarray}
where $\tau^{\mu\nu}$ is the surface stress tensor
from the localized source, giving the jump in ${\cal J}^{\mu\nu}$.
The surface action also gives rise to
the source $\Delta^\phi$ for the jump in ${\cal J}^\phi$.
From Eqs.~(\ref{delS}) and~(\ref{delSSigma}), we obtain~\cite{Padilla:2012ze}
\begin{eqnarray}
\left[{\cal J}^{\mu\nu}\right]^+_-
=-\frac{1}{2}\tau^{\mu\nu},\label{junction-cond}
\end{eqnarray}
and
\begin{eqnarray}
\left[{\cal J}^\phi\right]^+_- = \Delta^\phi,\label{junction-condi-phi}
\end{eqnarray}
where
$[(\cdots)]^+_-:=(\cdots)|_{\Sigma_+}-(\cdots)|_{\Sigma_-}$.
The above junction conditions together with the continuity
$[\gamma_{\mu\nu}]_-^+=0$ and $[\phi]_-^+=0$
determine how the metric and the scalar field
are matched across the surface $\Sigma$.
It is now clear from those conditions that
the first time derivatives of the metric and $\phi$
can be discontinuous, and hence the second time derivatives can be singular
at $\Sigma$.


\section{Cosmological matching conditions}

We consider a slightly perturbed universe whose metric
is given by
\begin{eqnarray}
\D s^2&=&-(1+2A)\D t^2+2B_i\D t\D x^i
\nonumber\\&&
+a^2\left[(1-2\psi)\delta_{ij}+2E_{ij}+h_{ij}\right]\D x^i\D x^j,
\end{eqnarray}
where $A$ and $\psi$ are scalar perturbations,
$h_{ij}$ is a traceless and transverse tensor perturbation,
and $B_i$ and $E_{ij}$ are decomposed into scalar and transverse vector parts as
\begin{eqnarray}
B_i=\partial_iB+B_i^{\rm V},
\quad
E_{ij}=\partial_i\partial_jE+\partial_{(i}E_{j)}^{\rm V}.
\end{eqnarray}
The scalar field also has a homogeneous
part and a small inhomogeneous perturbation as
$\phi(t,\mathbf{x})=\bar\phi(t)+\delta\phi(t,\mathbf{x})$.
We will omit the bar on the homogeneous part when there is no worry about confusion.

Let the matching surface be specified by $q(t,\mathbf{x})=0$.
This equation can be decomposed as $\bar q(t)+\delta q(t, \mathbf{x})=0$.
The cosmological matching conditions on this hypersurface are derived by calculating
${\cal J}_{i}^{\;j}=\overline{\cal J}_{i}^{\;j}(t)+\delta{\cal J}_{i}^{\;j}(t,\mathbf{x})$
and ${\cal J}^\phi=\overline{\cal J}^\phi(t)+\delta{\cal J}^\phi(t,\mathbf{x})$.
By using the temporal gauge transformation $t\to \tilde t=t+\xi^0$,
one can move to the {\em uniform $q$ gauge}, i.e., the coordinate system satisfying
\begin{eqnarray}
\delta q\to\widetilde{\delta q}=\delta q-\dot q\xi^0 =0.
\end{eqnarray}
Then, the matching surface is determined simply by the equation $\bar q(\tilde t)=0$,
or, equivalently, $\tilde t=$ const $=:t_*$.
Although the choice of the temporal gauge has no relevance to
the matching conditions for the homogeneous background
and tensor and vector perturbations,
this coordinate system is convenient for the computation of $\delta{\cal J}^\phi$ and
the scalar part of $\delta {\cal J}_{i}^{\;j}$.
A particular example of $q$ is $q=\phi(t,\mathbf{x})-\phi_*$,
where $\phi_*$ is some constant.
Another example is $q=\rho(t,\mathbf{x})-\rho_*$.

\subsection{Matching conditions for the homogeneous background}

Let us first consider
the matching conditions for a homogeneous and isotropic background.
The homogeneous part of ${\cal J}_{ij}$ is of the form
$\overline{\cal J}_{i}^{\;j}=(1/3)\overline{\cal J}\delta_{i}^{\;j}$,
where
\begin{eqnarray}
\frac{1}{3}\overline{\cal J}(\phi, \dot\phi, H)&=&-\frac{1}{2}f_3+2G_4H-4HXG_{4X}+\dot\phi G_{4\phi}
\nonumber\\&&
-H^2X\dot\phi G_{5X}+2HXG_{5\phi},\label{JC-back}
\end{eqnarray}
with
\begin{eqnarray}
f_3(\phi, X):=\int^X_0\sqrt{2u}G_{3u}(\phi, u)\D u.
\end{eqnarray}
Assuming that there are no localized sources on $\Sigma$,
the matching conditions for the background are given by
$[a]^+_-=0$ and
\begin{eqnarray}
\left[\,\overline{\cal J}(\phi, \dot\phi, H)\,\right]^+_-=0.
\label{bgjunc}
\end{eqnarray}
In general relativity
Eq.~(\ref{bgjunc}) reduces to the standard matching condition $[H]_-^+=0$.

The same condition can be derived by integrating the background equation
${\cal P}=-p$ (see Appendix)
from $t=t_*-\epsilon$ to $t=t_*+\epsilon$.
Isolating the second time derivatives
and denoting them with the subscript $\bullet\bullet$, one gets
\begin{eqnarray}
{\cal P}_{\bullet\bullet}&=&
\left(-2X G_{3X}+\cdots\right)\ddot\phi + \left(4G_{4}+\cdots\right)\dot H
\nonumber\\&=&
\left(\frac{2}{3}\partial_t\overline{\cal J}\right)_{\bullet\bullet},
\end{eqnarray}
which implies Eq.~(\ref{bgjunc}).

It is worth emphasizing that
even if $G_4=$ const and $G_5=0$, i.e., even if $\phi$ is minimally coupled to gravity,
$G_{3X}$ gives rise to a nonstandard term $f_3$ in the junction condition~(\ref{JC-back}).
This is because the gravitational field equations contain second derivatives of $\phi$
in the presence of $G_{3X}$.

Similarly, it is straightforward to get
\begin{eqnarray}
-\overline{{\cal J}}^\phi(\phi,\dot\phi, H) =J+f_{3\phi}-6HG_{4\phi},
\end{eqnarray}
where
\begin{eqnarray}
J&:=&\dot\phi G_{2X}+6HXG_{3X}-2\dot\phi G_{3\phi}
\nonumber\\&&
+6H^2\dot\phi\left(G_{4X}+2XG_{4XX}\right)-12HXG_{4\phi X}
\nonumber\\&&
+2H^3X\left(3G_{5X}+2XG_{5XX}\right)
\nonumber\\&&
-6H^2\dot\phi\left(G_{5\phi}+XG_{5\phi X}\right).
\end{eqnarray}
In the absence of a localized source, we obtain the scalar-field matching condition
\begin{eqnarray}
\bigl[\,\overline{{\cal J}}^\phi(\phi, \dot\phi, H)\,\bigr]^+_-=0,\label{match-scalar}
\end{eqnarray}
with the continuity $[\phi]^+_-=0$.  In general relativity with a
scalar field whose kinetic term is canonical, we have $\overline{\cal
J}^\phi = -\dot\phi$.  The same equation as Eq.~(\ref{match-scalar}) can
be derived as well by integrating the scalar-field equation of motion
(see Appendix) from $t=t_*-\epsilon$ to $t=t_*+\epsilon$, noting that
the second derivatives in the scalar-field equation are given by
\begin{eqnarray}
\left(\dot J-P_\phi\right)_{\bullet\bullet}
&=&\left(G_{2X}+\cdots\right)\ddot\phi+\left(6XG_{3X}+\cdots\right)\dot H,
\nonumber\\
&=&\left(-\partial_t\overline{\cal J}^\phi\right)_{\bullet\bullet}.
\end{eqnarray}

The matching conditions~(\ref{bgjunc}), (\ref{match-scalar}), and $[\phi]^+_-=0$
admit the solution satisfying the same conditions as in general relativity:
\begin{eqnarray}
\bigl[\,H\,\bigr]^+_- = 0,
\quad
\bigl[\,\dot\phi\,\bigr]^+_- = 0.\label{match-GR}
\end{eqnarray}
The second derivatives, $\dot H$ and $\ddot\phi$, can however be discontinuous
(but not singular) across $\Sigma$.
Obviously, $H^+$ and $\dot\phi^+$
determined from Eq.~(\ref{match-GR}) satisfy the Hamiltonian constraint,
${\cal E}(\phi^+,\dot\phi^+,H^+)=-\rho^+$.
There could be other nontrivial solutions, $H^+\neq H^-$, $\dot\phi^+\neq\dot\phi^-$,
to the matching conditions~(\ref{bgjunc})
and~(\ref{match-scalar}).
However, in contrast to the trivial solution~(\ref{match-GR}),
such solutions never satisfy the Hamiltonian constraint.
Thus, in the absence of any localized sources,
the first derivatives, $H$ and $\dot\phi$, must be continuous across $\Sigma$,
i.e., there is no essential modification
compared with the result of general relativity.
This is indeed the case if
the matter equation of state undergoes a sudden transition,
$p=p_-(\rho)\;\to\;p_+(\rho)$, at some $\rho=\rho_*=$ const hypersurface.
Another example
is the model where the nonsingular bounce is caused by
some scalar-field dynamics: in the scenario of~\cite{Cai:2013kja},
$H$ and $\dot \phi$ are continuous while $\dot H$ and $\ddot\phi$
can be approximated to be discontinuous at the beginning and end of the
bounce phase.

\begin{figure}[t]
  \begin{center}
    \includegraphics[keepaspectratio=true,height=100mm]{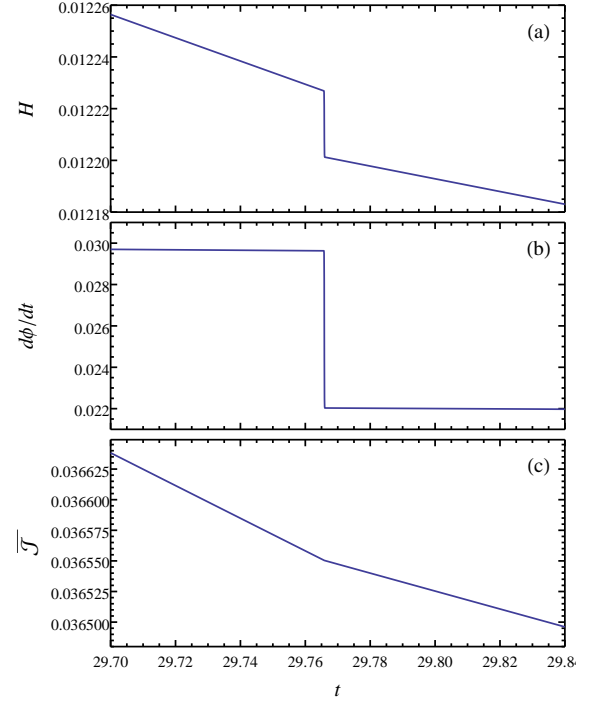}
  \end{center}
  \caption{Numerical example of sudden slow down of $\phi$
  caused by a steep hyperbolic tangent potential: (a) - $H$; (b) - $\dot\phi$;
  (c) - $\overline{\cal J}$.
  }%
  \label{fig:num.eps}
\end{figure}

To see the situation where the matching conditions
do not reduce simply to Eq.~(\ref{match-GR}),
let us investigate the model
with a step-like potential for the scalar field,
$G_2\supset -V_0\theta(\phi -\phi_*)$.
In deriving the above scalar matching condition,
we have implicitly assumed that the singular part
in the scalar-field equation of motion comes only from
the second derivatives $\dot H$ and $\ddot \phi$.
However, variation with respect to $\phi$ now gives
$\delta_\phi G_2\supset -V_0\delta(\phi-\phi_*)\delta\phi$,
leading to a non-vanishing localized
source of the jump in the right hand side of Eq.~(\ref{match-scalar}).
Equivalently, one 
can collect
the singular part of the scalar-field equation of motion,
\begin{eqnarray}
\left(\dot J-P_\phi\right)_{\rm sing}
=
\left(-\partial_t\overline{\cal J}^\phi\right)_{\bullet\bullet }
+V_0\delta(\phi-\phi_*),
\end{eqnarray}
to see that
the scalar matching condition in the form~(\ref{match-scalar})
cannot be used due to
the extra singular contribution $V_0\delta(\phi-\phi_*)$.
In this case, both $H$ and $\dot\phi$ are discontinuous in general,
but $\overline{\cal J}$ is continuous.
In the genesis scenario which will be discussed in the
next section~\cite{Creminelli:2010ba, Creminelli:2012my, Hinterbichler:2012fr, Hinterbichler:2012yn},
such a step in the potential will cause
an instantaneous change in $\dot\phi$ to end the genesis phase.

We present our numerical result in Fig.~\ref{fig:num.eps} corresponding to
the situation in which $\phi$ suddenly slows down.
As a simple example containing only $G_2$ and $G_3$
plus the Einstein-Hilbert term~\cite{Deffayet:2010qz,Kobayashi:2010cm},
the Lagrangian
\begin{eqnarray}
{\cal L}=\frac{R}{2}+X-cX\Box\phi -\frac{V_0}{2}\left[1+\tanh(\xi \phi)\right]
\end{eqnarray}
with $c, V_0=$ const, and $\xi \gg 1$
is employed for the numerical calculation to mimic the case of the step-like potential.
It can be seen that
$H$ and $\dot\phi$ experience a sharp jump,
but the matching condition~(\ref{bgjunc}) still holds.

\subsection{Matching conditions for cosmological perturbations}


We are now in position to consider matching of cosmological perturbations.
The matching conditions for scalar, vector, and tensor modes
can be studied separately.
\vspace{3mm}

{\bf (i)\;Tensor perturbations}

The transverse and traceless part of $\widetilde{\delta{\cal J}}_i^{\;j}$ is
\begin{eqnarray}
\widetilde{\delta {\cal J}}_i^{\;j}=-\frac{1}{4}\left({\cal G}_T\dot h_i^{\;j}+\frac{f_5}{a^2}
\partial^2h_i^{\;j}\right),
\end{eqnarray}
where
\begin{eqnarray}
{\cal G}_T:=2\left[G_4-2XG_{4X}-X\left(H\dot\phi G_{5X}-G_{5\phi}\right)\right],
\end{eqnarray}
and $f_5$ is defined similarly to $f_3$ as
\begin{eqnarray}
f_5(\phi, X):=\int^X_0\sqrt{2u}G_{5u}(\phi, u)\D u.
\end{eqnarray}
Thus, the matching conditions for the tensor perturbations are given by
\begin{eqnarray}
\left[h_{ij}\right]_-^+ =0,\quad
\left[{\cal G}_T \dot h_{ij}+\frac{f_5}{a^2}\partial^2h_{ij} \right]^+_-=0.
\label{Tensor-match}
\end{eqnarray}
Since the tensor perturbations are subject to a second-order differential equation,
the above two conditions are enough to determine their evolution after the transition.

The matching conditions~(\ref{Tensor-match}) can further be simplified
if $H$ and $\dot\phi$ are continuous across $\Sigma$. For such a background,
${\cal G}_T$ and $f_5$ are continuous and hence the matching conditions~(\ref{Tensor-match})
reduce to the same ones as in general relativity:
\begin{eqnarray}
\left[h_{ij}\right]_-^+ =0,\quad\bigl[\dot h_{ij}\bigr]_-^+ =0.
\end{eqnarray}

\vspace{3mm}

{\bf (ii)\; Vector perturbations}

The vector part of $\widetilde{\delta{\cal J}}_i^{\;j}$ is of the form
\begin{eqnarray}
\widetilde{\delta {\cal J}}_i^{\;j}=\delta^{jk}\partial_{(i}\delta{\cal J}_{k)}^{\rm V},
\quad
\delta{\cal J}_i^{\rm V}
:=\frac{{\cal G}_T}{2}\left(\frac{B_i^{\rm V}}{a^2}-\dot E_i^{\rm V}\right).
\end{eqnarray}
Note that $B_i^{\rm V}/a^2-\dot E_i^{\rm V}$ is the gauge-invariant combination.
The matching condition for the vector perturbations is therefore given in
any coordinates by
\begin{eqnarray}
\left[{\cal G}_T\left(\frac{B_i^{\rm V}}{a^2}-\dot E_i^{\rm V}\right)\right]^+_-=0.
\label{vector-junction}
\end{eqnarray}
The continuity of the induced metric, $[E_i^{\rm V}]^+_-=0$,
can always be satisfied by choosing the spatial coordinates appropriately.
Since the vector perturbations are governed by a first-order differential equation,
Eq.~(\ref{vector-junction}) completely fixes the integration constant at the transition.
If $H$ and $\dot\phi$ (and hence ${\cal G}_T$) are continuous,
Eq.~(\ref{vector-junction}) reduces to the same
matching condition as in general relativity.

\vspace{3mm}

{\bf (iii)\;Scalar perturbations}

As mentioned above, we will work in the uniform $q$ gauge,
$\delta q=q(t,\mathbf{x})-\bar q(t)=0$.
In this gauge,
the continuity implies that
\begin{eqnarray}
\bigl[\,\tilde\psi\,\bigr]_-^+=0,
\quad
\bigl[\,\tilde E\,\bigr]_-^+=0,
\quad
\bigl[\,\widetilde{\delta\phi}\,\bigr]_-^+=0,\label{scont}
\end{eqnarray}
where the second condition can always be satisfied
by choosing appropriately the spatial coordinates.
The junction conditions are derived from
\begin{widetext}
\begin{eqnarray}
\widetilde{\delta{\cal J}}_i^{\;j}&=&
\left[
-\left({\cal G}_T\dot{\tilde\psi }+\Theta \tilde A\right)
+\frac{\Theta -H{\cal G}_T}{\dot\phi}\dot{\widetilde{\delta\phi}}
+\frac{1}{3}\frac{\partial \overline{{\cal J}}}{\partial\phi}\widetilde{\delta\phi}
\right]\delta_i^{\;j}
+\frac{1}{2a^2}
\left(\partial_i\partial^j-\delta_i^{\;j}\partial^2\right)\left[
-{\cal G}_T
\tilde\sigma
-2W\widetilde{\delta\phi}
+f_5
\tilde \psi\right],
\\
\widetilde{\delta{\cal J}}^\phi&=&
-\frac{2}{\dot\phi}\left[
3\left(\Theta-H{\cal G}_T\right)\dot{\tilde\psi}
-\left(\Sigma+3H\Theta\right)\tilde A-\frac{\Theta-H{\cal G}_T}{a^2}\partial^2
\tilde\sigma\right]+
\frac{\partial \overline{{\cal J}}^\phi}{\partial\phi}\widetilde{\delta\phi}-
\frac{\Sigma+6H\Theta-3H^2{\cal G}_T}{X}
\dot{\widetilde{\delta\phi}}
\nonumber\\&&
+\frac{Z}{a^2}\partial^2\widetilde{\delta\phi}-\frac{4W}{a^2}\partial^2\tilde\psi,
\end{eqnarray}
\end{widetext}
where we defined the shear as
\begin{eqnarray}
\sigma:=a^2 \dot E-B,
\end{eqnarray}
and
\begin{eqnarray}
W&:=&\dot\phi G_{4X}+HXG_{5X}-\dot\phi G_{5\phi}+\frac{f_{5\phi}}{2},
\\
Z&:=&
\dot\phi G_{3X}+F_{3\widetilde X}-4F_{4\phi \widetilde X}+4HG_{4X}
\nonumber\\&&
+8HXG_{4XX}-2\dot\phi G_{4\phi X}+2\dot\phi H^2G_{5X}
\nonumber\\&&
+2\dot\phi XH^2G_{5XX}
-4HG_{5\phi}-4XHG_{5\phi X}.
\end{eqnarray}
The above equations yield the uniform $q$ gauge
expression of the matching conditions. However,
using the following formulas,
\begin{eqnarray}
&&\tilde A =A-\partial_t\left(\frac{\delta q}{\dot q}\right),
\quad
\tilde \psi =\psi+H\frac{\delta q}{\dot q},
\nonumber\\&&
\tilde\sigma=\sigma -\frac{\delta q}{\dot q},
\quad
\widetilde{\delta\phi}=\delta\phi-\dot\phi\frac{\delta q}{\dot q},
\label{gtrans}
\end{eqnarray}
one can
undo the gauge fixing to move from one gauge to the other.


Let us first consider the case where
the equation of state of matter experiences a sudden jump,
$p=p_-(\rho)\to p_+(\rho)$, at the time when $\rho=\rho_*=$ const,
so that
$\Sigma$ is determined by the equation
\begin{eqnarray}
q(t,\mathbf{x})=\rho(t,\mathbf{x})-\rho_*=0.
\end{eqnarray}
We assume that the localized source is absent on $\Sigma$,
and so can use the matching conditions for the background in the form
$[\overline{\cal J}]^+_-=[\overline{\cal J}^\phi]^+_-=0$.
From the discussion in the previous subsection, it turns out in the end that
$H$ and $\dot\phi$ are continuous.
From Eq.~(\ref{gtrans}) we see that the continuity~(\ref{scont})
can be written in an arbitrary gauge as
\begin{eqnarray}
\left[\psi + H\frac{\delta\rho}{\dot\rho}\right]_-^+&=&0,\label{mc1}
\\
\left[\delta\phi-\dot\phi\frac{\delta\rho}{\dot\rho}\right]_-^+&=&0.\label{mc2}
\end{eqnarray}
The trace and traceless parts of the equations $[\widetilde{\delta{\cal J}}_i^{\;j}]=0$
reduce in an arbitrary gauge to
\begin{eqnarray}
&&\biggl[
{\cal G}_T\dot\psi+\Theta A-\frac{\Theta-H{\cal G}_T}{\dot\phi}\dot{\delta\phi}
\nonumber\\&&\quad\qquad
+\left({\cal G}_T\dot H
+\frac{\Theta-H{\cal G}_T}{\dot\phi}\ddot\phi\right)\frac{\delta\rho}{\dot\rho}
\biggr]_-^+
=0,\label{mc4}
\end{eqnarray}
and
\begin{eqnarray}
\left[\sigma-\frac{\delta\rho}{\dot\rho}\right]^+_-=0,\label{mc3}
\end{eqnarray}
respectively.
Using Eq.~(\ref{mc4}), the matching condition $[\widetilde{\delta{\cal J}}^\phi]^+_-=0$
in an arbitrary gauge reads
\begin{eqnarray}
&&\biggl[-3\Theta\dot\psi+\Sigma A-\frac{1}{\dot\phi}\left(\Sigma+3H\Theta\right)\dot{\delta\phi}
\nonumber\\&&\quad\qquad
+\left(-3\Theta\dot H+\frac{\Sigma +3H\Theta}{\dot\phi}\ddot\phi\right)\frac{\delta\rho}{\dot\rho}
\biggr]_-^+
=0.\label{mcjp}
\end{eqnarray}
The two conditions~(\ref{mc4}) and~(\ref{mcjp})
can be rearranged to give
\begin{eqnarray}
\left[\dot\psi+\frac{H}{\dot\phi}\dot{\delta\phi}+
\left(\dot H-H\frac{\ddot\phi}{\dot\phi}\right)\frac{\delta\rho}{\dot\rho}\right]^+_-
&=&0,\label{jc1}
\\
\left[A-\frac{\dot{\delta\phi}}{\dot\phi}
+\frac{\ddot\phi}{\dot\phi}\frac{\delta\rho}{\dot\rho}\right]_-^+&=&0.\label{jc2}
\end{eqnarray}
Interestingly, these matching conditions are independent of
the concrete form of $G_i(\phi, X)$,
and hence are the same as those in general relativity
with a conventional scalar field.
Note, however, that the matching procedure requires the use of
the constraint equations presented in Appendix,
which depend on the concrete form of $G_i(\phi, X)$.

Having thus obtained the matching conditions in an arbitrary gauge,
let us see how one can consistently determine the perturbation variables
at $t=t_*+\epsilon$ in the unitary gauge ($\delta\phi =0$).
In this gauge, Eq.~(\ref{mc2}) reads $[\delta\rho_u/\dot\rho]^+_- =0$,
and then Eq.~(\ref{mc1}) implies that
\begin{eqnarray}
[{\cal R}]^+_- = 0,
\end{eqnarray}
where ${\cal R}$ is the curvature perturbation in the unitary gauge,
\begin{eqnarray}
{\cal R}:=\psi+H\frac{\delta\phi}{\dot\phi}.
\end{eqnarray}
Here and hereafter the subscript $u$ refers to the unitary gauge variable.
Equation~(\ref{mc3}) simply becomes $[\sigma_u]^+_- =0$.
Equations~(\ref{jc1}) and~(\ref{jc2}) can be
used to determine $\dot{\cal R}$ and $A_u$
at $t=t_*+\epsilon$.
The Hamiltonian constraint is consistent with Eq.~(\ref{mcjp}),
while the momentum constraint is used to fix the velocity perturbation $\delta u_u$.
Thus, all the perturbation variables
at $t=t_*+\epsilon$ can be determined.

The matching procedure
in the Newtonian gauge ($\sigma=0$)
is slightly different from that in the unitary gauge.
In the Newtonian gauge,
Eq.~(\ref{mc3}) reads $[\delta\rho_N/\dot\rho]_-^+=0$,
where the subscript $N$ stands for the Newtonian gauge variable.
In terms of the metric potentials in the Newtonian gauge,
\begin{eqnarray}
\Phi:=A-\dot\sigma,\quad
\Psi:=\psi+H\sigma,
\end{eqnarray}
Eqs.~(\ref{mc1}) and~(\ref{mc2}) are rewritten as
\begin{eqnarray}
[\Psi]^+_- =0,\quad[\delta\phi_N]_-^+=0,
\end{eqnarray}
while Eqs.~(\ref{jc1}) and~(\ref{jc2}) yields
the two relations among $\dot\Psi$, $\Phi$, and $\dot{\delta\phi}_N$.
We then invoke the traceless part of the $(i,j)$ components of the field equations,
\begin{eqnarray}
{\cal G}_T\Phi-{\cal F}_T\Psi+
\frac{\dot{\cal G}_T+H({\cal G}_T-{\cal F}_T)}{\dot\phi}\delta\phi_N=0,
\label{trac}
\end{eqnarray}
to remove $\Phi$, and thus determine $\dot\Psi$ and $\dot{\delta\phi}_N$
at $t=t_*+\epsilon$.

The next example we would like to study is
the transition that occurs when
$\phi$ reaches some value $\phi_*$:
\begin{eqnarray}
q(t,\mathbf{x})=\phi(t,\mathbf{x})-\phi_*.
\end{eqnarray}
In the previous example of $q=\rho(t,\mathbf{x})-\rho_*$,
we considered the case where $H$ and $\dot\phi$ are continuous.
In present example, however, we allow for discontinuous $H$ and $\dot\phi$,
because such a situation can easily be realized at the moment when
$\phi(t,\mathbf{x})$ passes a step in the potential at $\phi_*$,
as already demonstrated.
In this case, it is convenient to stay in the uniform $q$ gauge
since it coincides with the uniform $\phi$ gauge.
Then, the curvature perturbation on uniform $\phi$
hypersurfaces is given by
${\cal R}=\psi+H\delta\phi/\dot\phi = \tilde\psi$, and
the matching conditions
$[\tilde\psi]^+_-=0$ and
$[\widetilde{\delta{\cal J}}_i^{\;j}]^+_-=0$ reduce to
\begin{eqnarray}
&&\bigl[{\cal R}\bigr]_-^+=0,\label{j1}
\\
&&\bigl[{\cal G}_T\dot{\cal R}+\Theta\tilde A\bigr]_-^+=0,\label{j2}
\\
&&\bigl[{\cal G}_T\tilde\sigma -f_5{\cal R}\bigr]_-^+=0.\label{j3}
\end{eqnarray}
Let us first assume for simplicity that usual matter is absent.
Equation~(\ref{j2}) automatically holds
thanks to the momentum constraints.
Combining the Hamiltonian and momentum constraints,
we find
\begin{eqnarray}
{\cal G}_S\dot {\cal R}-\frac{1}{a^2}\frac{{\cal G}_T^2}{\Theta}\partial^2{\cal R}
-\frac{{\cal G}_T}{a^2}\partial^2\tilde\sigma =0.
\end{eqnarray}
The matching condition~(\ref{j3}) then reads
\begin{eqnarray}
\left[
{\cal G}_S\dot{\cal R}-\frac{1}{a^2}\left(\frac{{\cal G}_T^2}{\Theta}+f_5\right)\partial^2
{\cal R}\right]^+_- =0.
\label{j4}
\end{eqnarray}
Using Eqs.~(\ref{j1}) and~(\ref{j4}) one can do the matching of ${\cal R}$
and $\dot{\cal R}$.
In the case where $\dot \phi$ and $H$ are continuous,
the latter condition is simplified to $[\dot{\cal R}]_-^+=0$.
However, if the second derivatives diverge and hence $\dot\phi$ and $H$
are discontinuous, one must employ the full equation~(\ref{j4}).

In the presence of usual matter, the matching condition~(\ref{j4})
is modified as
\begin{eqnarray}
&&\left[
{\cal G}_S\dot{\cal R}-\frac{1}{a^2}\left(\frac{{\cal G}_T^2}{\Theta}+f_5\right)\partial^2
{\cal R}\right]^+_- 
\nonumber\\&& \quad +
\left[\frac{{\cal G}_T}{2\Theta}\left(\widetilde{\delta \rho}
+\frac{\Sigma}{\Theta}\left(\rho+p\right)\widetilde{\delta u}\right)\right]_-^+
=0,\label{j5mod}
\end{eqnarray}
while Eq.~(\ref{j1}) remains unchanged.
Since the continuity and Euler equations for matter do not contain
second derivatives of the metric, all the matter-related quantities
are continuous across the matching surface specified by $q=\phi(t,\mathbf{x})-\phi_*=0$.
If the transition is such that $\dot\phi$ and $H$ are continuous,
then Eq.~(\ref{j5mod}) implies that we are
still allowed to use the condition $[\dot{\cal R}]_-^+=0$.

\section{Genesis models from Horndeski's theory}

In this section, we demonstrate how the matching conditions
are used at the transition from the galilean genesis phase to
the standard radiation-dominated Universe.
This is probably the most illustrative example
because stable galilean genesis is realized
thanks to the terms ${\cal L}_i$ with $i\ge 3$,
which give rise to the new boundary terms.
The matching procedure has been carried out
in specific examples of galilean genesis in
Refs.~\cite{Creminelli:2012my, Hinterbichler:2012yn}.
We will extend those previous models
and present a unified analysis of the theory admitting galilean genesis.
To do so,
we generalize the Lagrangian of Ref.~\cite{Rubakov:2013kaa} and
study a subclass of Horndeski's theory defined by
\begin{eqnarray}
&&G_2=e^{4\lambda\phi}g_2(Y),
\quad
G_3=e^{2\lambda\phi} g_3(Y),
\nonumber\\
&&G_4=\frac{M^2_{\rm Pl}}{2}+e^{2\lambda\phi}g_4(Y),
\quad
G_5=e^{-2\lambda\phi}g_5(Y),\label{gen:Lag}
\end{eqnarray}
where each $g_i\;(i=2,3,4,5)$ is a function of
\begin{eqnarray}
Y:=e^{-2\lambda\phi} X,
\end{eqnarray}
and $\lambda$ and $M_{\rm Pl}$ are constants.
We assume that $g_4(0)=0$.

Let us look for a solution of the form
\begin{eqnarray}
e^{\lambda\phi}\simeq \frac{1}{\lambda\sqrt{2Y_0}}\frac{1}{(-t)},
\quad
H\simeq\frac{h_0}{(-t)^3}
\quad (-\infty <t<0),\label{gen:back}
\end{eqnarray}
where $Y_0$ and $h_0$ are positive constants.
Note that $Y\simeq Y_0$ for this background.
Equation~(\ref{gen:back})
should be regarded as an approximate solution valid for $|t|\gg \sqrt{h_0}$,
and in this section we only consider the case where this approximation is good.
The spacetime is close to Minkowski for $|t|\gg \sqrt{h_0}$ and expands as
$a\simeq 1+h_0(-t)^{-2}/2$.
Since $\dot H=3h_0(-t)^{-4}>0$, one can interpret this solution to be
NEC violating.
The above solution is essential for the galilean genesis
scenario~\cite{Creminelli:2010ba, Creminelli:2012my, Hinterbichler:2012fr, Hinterbichler:2012yn}.
The Lagrangian defined by Eq.~(\ref{gen:Lag}) contains
different models of galilean genesis as specific cases,
and allows us to
study the genesis scenario in a unified manner.\footnote{The DBI conformal
galileons used in Ref.~\cite{Hinterbichler:2012yn} can be reproduced
from $G_5=\tilde g_5(Y)$, rather than $G_5=e^{-2\lambda\phi} g_5(Y)$.
However, for the genesis background~(\ref{gen:back}),
the contribution from $\tilde g_5(Y)$ is subleading for $|t|\gg\sqrt{h_0}$
compared to the other terms,
and hence has no effect on any equations.}

The background equations read
\begin{eqnarray}
{\cal E}&\simeq&
2XG_{2X}-G_2-2X G_{3\phi}
\nonumber\\
&=& e^{4\lambda\phi}\hat\rho(Y_0)=0,\label{gen:eq1}
\\
{\cal P}&\simeq &
4\left(G_4+XG_{5\phi}\right)\dot H+G_2-2X\left(G_{3\phi}+\ddot\phi G_{3X}\right)
\nonumber\\&&+2\ddot\phi G_{4\phi}+4XG_{4\phi\phi}+4X\ddot\phi G_{4\phi X}
\nonumber\\&&+4HX\dot XG_{5\phi X}+4H\dot X G_{5\phi}+4HX\dot\phi G_{5\phi\phi}
\nonumber\\&=&2{\cal G}(Y_0)\dot H+e^{4\lambda\phi}\hat p(Y_0)=0,\label{gen:eq2}
\end{eqnarray}
where
\begin{eqnarray}
\hat\rho(Y)&:=&2Y g_2'-g_2-4\lambda Y\left(g_3-Yg_3'\right),
\\
\hat p(Y)&:=&g_2-4\lambda Yg_3+24\lambda^2 Y\left(g_4-Yg_4'\right),
\\
{\cal G}(Y)&:=&M^2_{\rm Pl}-4\lambda Y\left(g_5+Yg_5'\right),
\end{eqnarray}
and a prime stands for differentiation with respect to $Y$. The constant
$Y_0$ is determined as a positive root of
\begin{eqnarray}
\hat\rho(Y_0)=0,\label{co1}
\end{eqnarray}
and then
$h_0$ is determined from Eq.~(\ref{gen:eq2}) as
\begin{eqnarray}
h_0= -\frac{1}{24\lambda^4}\frac{\hat p(Y_0)}{Y_0^2{\cal G}(Y_0)}.\label{h0determined}
\end{eqnarray}
As will be seen shortly, this background is stable for ${\cal G}(Y_0)>0$.
Hence, the above NEC violating solution is possible provided that
\begin{eqnarray}
\hat p(Y_0) <0.\label{co2}
\end{eqnarray}

For tensor perturbations, it is straightforward to compute
\begin{eqnarray}
{\cal G}_T \simeq {\cal G}(Y_0),
\quad
{\cal F}_T \simeq M^2_{\rm Pl}+4\lambda Y_0 g_5(Y_0),
\end{eqnarray}
and therefore the background is stable against tensor perturbations if
\begin{eqnarray}
&&{\cal G}(Y_0)>0,\label{co3}\\
&&M^2_{\rm Pl}+4\lambda Y_0 g_5(Y_0)>0.\label{co4}
\end{eqnarray}
For scalar perturbations, we find
\begin{eqnarray}
{\cal G}_S\simeq\left(\frac{{\cal G}}{\Theta}\right)^2\Sigma,
\quad
{\cal F}_S\simeq\left(\frac{{\cal G}}{\Theta}\right)^2\left(-\dot\Theta\right),
\end{eqnarray}
where
\begin{eqnarray}
\Sigma&\simeq &e^{4\lambda\phi} Y_0\hat\rho'(Y_0),
\\
\Theta&\simeq&\left[{\cal G}(Y_0)+2Y_0{\cal G}'(Y_0)\right]H
\nonumber\\&&
+\frac{\dot\phi e^{2\lambda\phi}}{12\lambda Y_0}\left[2Y_0\hat p'(Y_0)-\hat p(Y_0)\right].
\label{gen-Theta}
\end{eqnarray}
From Eq.~(\ref{gen-Theta}) it is easy to show
\begin{eqnarray}
-\dot\Theta\simeq \left.2\dot H\left(-\hat p\right)\left(\frac{Y{\cal G}}{\hat p}\right)'\right|_{Y_0},
\end{eqnarray}
so that the background is stable against scalar perturbations if
\begin{eqnarray}
&&\hat\rho'(Y_0)>0,\label{co5}
\\
&&\left.\left(\frac{Y{\cal G}}{\hat p}\right)'\right|_{Y_0}>0.\label{co6}
\end{eqnarray}
Since ${\cal G}_S\propto(-t)^2$ and ${\cal F}_S\propto(-t)^2$,
the sound speed, $c_s=\sqrt{{\cal F}_S/{\cal G}_S}$,
stays constant during the genesis phase.

The genesis phase is supposed to be
followed by the standard radiation-dominated phase.
As in~\cite{Creminelli:2012my}, we consider the model in which
the transition is caused by sudden halt of the scalar field
due to some upward lift of its potential, $G_2\supset -V_0\theta(\phi-\phi_*)$.
Then, the second derivatives $\ddot\phi$ and $\dot H$
diverge at $t=t_*$.
We neglect the contribution from the scalar field to the expansion rate
in the radiation-dominated phase, assuming $\dot\phi_{\rm rad}\simeq 0$.
It is then found that
\begin{eqnarray}
\frac{1}{3}\overline{{\cal J}}_{\rm gen}
&\simeq&
{\cal G}(Y_0)H_{\rm gen}-\frac{e^{3\lambda\phi}}{2}\int_0^{Y_0}\sqrt{2y}g_3'(y)\D y
\nonumber\\&&
+2\lambda\dot\phi e^{2\lambda\phi}\left(g_4-Y_0g_4'\right),
\end{eqnarray}
in the genesis phase
and $(1/3)\overline{\cal J}_{\rm rad}\simeq M^2_{\rm Pl}H_{\rm rad}$
in the radiation-dominated phase.
The radiation-dominated universe is required to be expanding, $H_{\rm rad}>0$.
The matching condition $\overline{{\cal J}}_{\rm gen}-\overline{{\cal J}}_{\rm rad}=0$ therefore
reads $\overline{{\cal J}}_{\rm gen}>0$.
Using Eq.~(\ref{h0determined}), this condition can be written as
\begin{eqnarray}
-g_2-2\lambda Y_0g_3(Y_0)+3\lambda\sqrt{Y_0}
\int_0^{Y_0}\frac{g_3(y)}{\sqrt{y}}\D y>0.\label{gentorad}
\end{eqnarray}
Note that this can be derived without relying on
what the dominant component in the post-genesis phase is;
we only require that the post-genesis universe is just expanding.

We have thus arrived at the generic conclusion based on
the Lagrangian defined by Eq.~(\ref{gen:Lag}) without
specifying its further concrete form:
a consistent genesis scenario is realized
provided
$\hat\rho(Y_0)=0$ has a positive root and
Eqs.~(\ref{co2}),~(\ref{co3}),~(\ref{co4}),~(\ref{co5}),~(\ref{co6}),
and~(\ref{gentorad}) are satisfied.
It is easy to fulfill all of these conditions simultaneously
even in the simple Lagrangian
with~\cite{Creminelli:2010ba,Creminelli:2012my}
\begin{eqnarray}
g_2=-Y+c_2Y^2,\quad g_3=c_3Y,\quad g_4=g_5=0,
\end{eqnarray}
where $c_2$ and $c_3$ are some constants.
Indeed, all the requirements are satisfied for
$4\lambda c_3>c_2>0$.




Let us then investigate the matching of the perturbation variables.
On superhorizon scales, the general solution to the tensor perturbation equation
in Fourier space
is given by
\begin{eqnarray}
h_{ij}&=&C_k^{g-}
-\frac{C_k^{d-}}{{\cal G}(Y_0)}\int_t^{t_*}\frac{\D t'}{a^3(t')}
\quad(t<t_*),
\\
h_{ij}&=&C_k^{g+}
+\frac{C_{k}^{d+}}{\mpl^2}\int_{t_*}^t\frac{\D t'}{a^3(t')}
\quad(t>t_*),
\end{eqnarray}
where $C_{k}^{g\pm}$ and $C_{k}^{g\pm}$ are integration constants
that depend on the wavenumber $k$.
From the matching conditions~(\ref{Tensor-match}),
the integration constants in the post-genesis phase are determined as
\begin{eqnarray}
C_{k}^{g+}=C_{k}^{g-},\quad C_{k}^{d+}=C_k^{d-}.
\end{eqnarray}
Note, however, that tensor perturbations generated during
the genesis phase are observationally irrelevant, because
$a\sim 1$, ${\cal G}_T, {\cal F}_T\sim$ const, so that
the vacuum fluctuations of $h_{ij}$ are not amplified.

As for the scalar perturbations, it is most convenient to
use the curvature perturbation on uniform $\phi$ slices, ${\cal R}$.
The central quantity for the matching of ${\cal R}$ is ${\cal G}_S$,
because from the matching conditions we see that
${\cal R}$ and ${\cal G}_S\dot{\cal R}$ are continuous
on superhorizon scales. In the present case, ${\cal G}_S$ is of the form
${\cal G}_S={\cal A}(Y_0)(-t)^2$ for $t<t_*$, and
${\cal G}_S={\cal G}_S^+=$ const for $t>t_*$, where
the concrete expression for ${\cal A}(Y_0)$
is not so illuminating. The superhorizon solution for ${\cal R}$
is given by
\begin{eqnarray}
{\cal R}&=&C^-_k
-\frac{D^-_k}{{\cal A}(Y_0)}\int_t^{t^*}\frac{\D t'}{(-t')^2a^3(t')}
\quad(t<t_*),\label{R-before}
\\
{\cal R}&=&C^+_k
+\frac{D^+_k}{{\cal G}_S^+}\int_{t_*}^{t}\frac{\D t'}{a^3(t')}
\quad(t>t_*),\label{R-after}
\end{eqnarray}
where it follows from the matching conditions that
\begin{eqnarray}
C^+_k=C^-_k,\quad
D^+_k=D^-_k.
\end{eqnarray}
The second term in Eq.~(\ref{R-before})
is matched to the second one in Eq.~(\ref{R-after}),
i.e., the decaying mode in the post-genesis Universe.
Thus, we have ${\cal R}\simeq C^-_k$ at sufficiently late times.
Following the usual quantization procedure we determine
$|C^-_k|=(2\sqrt{c_sk{\cal A}(Y_0)}|t_*|)^{-1}\sim k^{-1/2}$,
and hence we cannot get the scale-invariant fluctuations.
See also Refs.~\cite{LevasseurPerreault:2011mw,Wang:2012bq}
for discussions about the spectrum of fluctuations from galilean genesis.

\section{Conclusions}

In this paper, we have obtained the matching conditions
for the homogeneous and isotropic Universe and for cosmological perturbations
in Horndeski's most general second-order scalar-tensor theory,
starting from the generalization of Israel's conditions~\cite{Padilla:2012ze}.
In the absence of any localized sources at the transition hypersurface,
we have shown that the first derivatives of the metric and the scalar field,
$H(t)$ and $\dot\phi(t)$, must be continuous as in the case of general relativity.
This is the case where the equation of state of matter undergoes a sharp change.
In the case where $\phi$ suddenly lose its velocity due to
a step in the potential, some combination $\overline{\cal J}$ of $H$ and $\dot\phi$,
defined in Eq.~(\ref{JC-back}), is continuous across the transition hypersurface.
For cosmological perturbations we have obtained the junction equations
that can be used in any gauge.

Horndeski's theory can accommodate exotic but stable cosmologies
such as galilean genesis~\cite{Creminelli:2010ba}.
The cosmological matching conditions we have presented in this paper
can be applied to such interesting scenarios.
To demonstrate this, we have developed a generic Lagrangian
admitting the genesis solution that starts expanding from
the Minkowski spacetime in the asymptotic past,
and presented the conditions under which
a stable genesis background is consistently joined to an expanding universe.

\acknowledgments
We thank Tony Padilla and Vishagan Sivanesan for their helpful correspondence.
This work was supported in part by JSPS Grant-in-Aid for Young
Scientists (B) No.~24740161 (T.K.), the Grant-in-Aid for Scientific
Research on Innovative Areas No.~24111706 (M.Y.), and the Grant-in-Aid
for Scientific Research No.~25287054 (M.Y.).  
The work of N.T. is supported in part by World Premier International 
Research Center Initiative (WPI Initiative), MEXT, Japan, 
and JSPS Grant-in-Aid for Scientific Research 25$\cdot$755. 

\appendix
\section{Field equations}

In this Appendix, we summarize the background and linearized equations
used in the main text.
More details can be found in Refs.~\cite{Kobayashi:2011nu, DeFelice:2011hq}.

\subsection{Background equations}\label{App:BGEQ}

The evolution of the homogeneous and isotropic background is
determined from
\begin{eqnarray}
{\cal E}=-\rho,\quad{\cal P}=-p,
\end{eqnarray}
where
\begin{eqnarray}
{\cal E}&:=&2XG_{2X}-G_2
+6X\dot\phi HG_{3X}-2XG_{3\phi}
\nonumber\cr&&
-6H^2G_4+24H^2X(G_{4X}+XG_{4XX})
\nonumber\cr&&
-12HX\dot\phi G_{4\phi X}-6H\dot\phi G_{4\phi }
\nonumber\cr
&&+2H^3X\dot\phi\left(5G_{5X}+2XG_{5XX}\right)
\nonumber\cr&&
-6H^2X\left(3G_{5\phi}+2XG_{5\phi X}\right),
\\
{\cal P}&=&G_2
-2X\left(G_{3\phi}+\ddot\phi G_{3X} \right)
\nonumber\cr
&&
+2\left(3H^2+2\dot H\right) G_4
-12 H^2 XG_{4X}-4H\dot X G_{4X}
\nonumber\cr&&
-8\dot HXG_{4X}-8HX\dot X G_{4XX}
+2\left(\ddot\phi+2H\dot\phi\right) G_{4\phi}
\nonumber\cr&&
+4XG_{4\phi\phi}
+4X\left(\ddot\phi-2H\dot\phi\right) G_{4\phi X}
\nonumber\cr&&
-2X\left(2H^3\dot\phi+2H\dot H\dot\phi+3H^2\ddot\phi\right)G_{5X}
\nonumber\cr&&
-4H^2X^2\ddot\phi G_{5XX}
+4HX\left(\dot X-HX\right)G_{5\phi X}
\nonumber\cr
&&
+2\left[2\left(HX\right){\bf \dot{}}+3H^2X\right]G_{5\phi}
+4HX\dot\phi G_{5\phi\phi},
\end{eqnarray}
and $\rho$ and $p$ are the energy density and pressure
of usual matter, respectively.
The first equation corresponds to the Friedmann equation
(the Hamiltonian constraint), and the second one
to the evolution equation containing the second derivatives of
the metric and the scalar field.
The equation of motion for $\phi$ is given by
\begin{eqnarray}
\dot J+3HJ&=&
P_\phi
\end{eqnarray}
where
\begin{eqnarray}
J&:=&\dot\phi G_{2X}+6HXG_{3X}-2\dot\phi G_{3\phi}
\nonumber\\&&
+6H^2\dot\phi\left(G_{4X}+2XG_{4XX}\right)-12HXG_{4\phi X}
\nonumber\\&&
+2H^3X\left(3G_{5X}+2XG_{5XX}\right)
\nonumber\\&&
-6H^2\dot\phi\left(G_{5\phi}+XG_{5\phi X}\right), 
\end{eqnarray}
and
\begin{eqnarray}
P_\phi&=&G_{2\phi}-2X\left(G_{3\phi\phi}+\ddot\phi G_{3\phi X}\right)
\nonumber\\&&
+6\left(2H^2+\dot H\right)G_{4\phi}
\nonumber\\&&
+6H\left(\dot X+2HX\right)G_{4\phi X}
\nonumber\\&&
-6H^2XG_{5\phi\phi}+2H^3X\dot\phi G_{5\phi X}.
\end{eqnarray}

\subsection{Linear perturbations}

In the main test we use the following equations for
scalar cosmological perturbations:
(i) the Hamiltonian constraint,
\begin{eqnarray}
&&-6\Theta\dot\psi
+2{\cal G}_T\frac{\partial^2\psi}{a^2}+2\Sigma A+\frac{2}{a^2}\Theta\partial^2 \sigma
-\frac{\partial{\cal E}}{\partial\phi}\delta\phi
\nonumber\\&&
-\frac{2}{\dot\phi}\left(\Sigma+3H\Theta\right)\dot{\delta\phi}
-\frac{2}{\dot\phi}\left(\Theta-H{\cal G}_T\right)\frac{\partial^2\delta\phi}{a^2}
=\delta\rho,\;\;\;
\end{eqnarray}
(ii) the momentum constraint,
\begin{eqnarray}
&&-2\left({\cal G}_T\dot\psi+\Theta A\right)
+\frac{2}{\dot\phi}\left(\Theta-H{\cal G}_T\right)\dot{\delta\phi}
+J\delta\phi
\nonumber\\&&
-2\Bigl(
HG_{4\phi}-\dot\phi G_{4\phi\phi}+4HXG_{4\phi X}
\nonumber\\&&
-2HXG_{5\phi\phi}+H^2X\dot\phi G_{5\phi X}
\Bigr)\delta\phi
=(\rho+p)\delta u,
\end{eqnarray}
and (iii) the traceless part of the $(i,j)$ equations,
\begin{eqnarray}
&&{\cal G}_TA-{\cal F}_T\psi
+\frac{\dot{\cal G}_T+H({\cal G}_T-{\cal F}_T)}{\dot\phi}\delta\phi
\nonumber\\&&
-{\cal G}_T\dot\sigma
-\left(\dot{\cal G}_T+H{\cal G}_T\right)\sigma=0,
\end{eqnarray}
where we have included the perturbations of the matter energy-momentum tensor:
$\delta T_0^{\;0}=-\delta\rho$, $\delta T_i^{\;0}=(\rho+p)\partial_i\delta u$,
and $\delta T_i^{\;j}=\delta p\delta_i^{\;j}$.
Energy-momentum conservation implies
\begin{eqnarray}
&&\partial_t\delta\rho+3H(\delta\rho+\delta p)-3(\rho+p)\dot\psi
+\frac{\rho+p}{a^2}\partial^2\left(\delta u+\sigma\right)=0,
\nonumber\\
&&\\
&&\partial_t\left[(\rho+p)\delta u\right]+3H(\rho+p)\delta u+(\rho+p)A+\delta p=0.
\end{eqnarray}
In the above we defined
\begin{eqnarray}
{\cal F}_T&:=&2\left[G_4
-X\left( \ddot\phi G_{5X}+G_{5\phi}\right)\right],
\\
{\cal G}_T&:=&2\left[G_4-2 XG_{4X}
-X\left(H\dot\phi G_{5X} -G_{5\phi}\right)\right],
\\
\Sigma&:=&XG_{2X}+2X^2G_{2XX}+12H\dot\phi XG_{3X}
\nonumber\cr&&
+6H\dot\phi X^2G_{3XX}
-2XG_{3\phi}-2X^2G_{3\phi X}-6H^2G_4
\nonumber\cr&&
+6\Bigl[H^2\left(7XG_{4X}+16X^2G_{4XX}+4X^3G_{4XXX}\right)
\nonumber\cr&&
-H\dot\phi\left(G_{4\phi}+5XG_{4\phi X}+2X^2G_{4\phi XX}\right)
\Bigr]
\nonumber\cr&&
+30H^3\dot\phi XG_{5X}+26H^3\dot\phi X^2G_{5XX}
\nonumber\cr&&
+4H^3\dot\phi X^3G_{5XXX}
-6H^2X\bigl(6G_{5\phi}
\nonumber\cr&&
+9XG_{5\phi X}+2 X^2G_{5\phi XX}\bigr),
\\
\Theta&:=&-\dot\phi XG_{3X}+
2HG_4-8HXG_{4X}
\nonumber\cr&&
-8HX^2G_{4XX}+\dot\phi G_{4\phi}+2X\dot\phi G_{4\phi X}
\nonumber\cr&&
-H^2\dot\phi\left(5XG_{5X}+2X^2G_{5XX}\right)
\nonumber\cr&&
+2HX\left(3G_{5\phi}+2XG_{5\phi X}\right).
\end{eqnarray}

The unitary gauge $\delta\phi=0$ is convenient
in the absence of usual matter. The evolution equation for the curvature perturbation
in the unitary gauge, ${\cal R}$, follows from the quadratic action
\begin{eqnarray}
S_{\cal R}^{(2)}=\int\D t\D^3 x\,a^3\left[
{\cal G}_S
\dot{\cal R}^2
-\frac{{\cal F}_S}{a^2}
(\partial{\cal R})^2
\right]\label{scalar2},
\end{eqnarray}
where
\begin{eqnarray}
{\cal F}_S&:=&\frac{1}{a}\frac{\D}{\D t}\left(\frac{a}{\Theta}{\cal G}_T^2\right)
-{\cal F}_T,
\\
{\cal G}_S&:=&\frac{\Sigma }{\Theta^2}{\cal G}_T^2+3{\cal G}_T.
\end{eqnarray}
Similarly, the quadratic action for the tensor perturbation is given by
\begin{eqnarray}
S_h^{(2)} =\frac{1}{8}\int\D t\D^3x\,a^3\left[
{\cal G}_T\dot h_{ij}^2-\frac{{\cal F}_T}{a^2}
(\partial h_{ij})^2\right]. \label{tensoraction}
\end{eqnarray}
From those actions we see that the scalar and tensor perturbations are stable if
\begin{eqnarray}
{\cal F}_T>0,\;\;
{\cal G}_T>0,\;\;
{\cal F}_S>0,\;\;
{\cal G}_S>0.
\end{eqnarray}



\end{document}